\documentclass{article}

\usepackage{arxiv}
\usepackage[T1]{fontenc}    
\usepackage{hyperref}       
\usepackage{url}            
\usepackage{booktabs}       
\usepackage{amsfonts}       
\usepackage{microtype}      
\usepackage{lipsum}
\usepackage{float}
\usepackage[utf8]{inputenc} 
\usepackage{graphicx}
\usepackage{amsmath, mathtools}
\usepackage{biblatex}
\makeatletter
\newcommand{\leqnomode}{\tagsleft@true\let\veqno\@@leqno}
\newcommand{\reqnomode}{\tagsleft@false\let\veqno\@@eqno}
\makeatother

\title{Automated quantum programming \protect\\ via reinforcement learning \protect\\ for combinatorial optimization}

\author{
  Keri A. McKiernan $^{\dagger}$\\
  Stanford University \\
  Stanford, CA 94306 \\
  \texttt{kmckiern@stanford.edu}\\
  \And
  Erik Davis \\
  Rigetti Computing \\
  Berkeley, CA 94710 \\
  \texttt{erik@rigetti.com} \\
  \And
  M. Sohaib Alam \\
  Rigetti Computing \\
  Berkeley, CA 94710 \\
  \texttt{sohaib@rigetti.com} \\
  \And
  Chad Rigetti \\
  Rigetti Computing \\
  Berkeley, CA 94710 \\
  \texttt{chad@rigetti.com} \\
}

\date{\today}

\begin{document}
\maketitle

\begin{abstract}
We develop a general method for incentive-based programming of hybrid quantum-classical computing systems using reinforcement learning, and apply this to solve combinatorial optimization problems on both simulated and real gate-based quantum computers. Relative to a set of randomly generated problem instances, agents trained through reinforcement learning techniques are capable of producing short quantum programs which generate high quality solutions on both types of quantum resources. We observe generalization to problems outside of the training set, as well as generalization from the simulated quantum resource to the physical quantum resource. 
\end{abstract}

\section{Introduction}
One of the earliest ambitions of artificial intelligence research was to consider ways of mechanizing the task of computer programming itself, via the automated synthesis of programs from high level specifications. There is rich literature on such techniques, including a range of meta-heuristic \cite{olmo2014swarm,boussaid2013survey} and machine learning approaches \cite{kant2018recent} (cf. \cite{gulwani2017program} for a recent, broad survey on program synthesis). Such ideas are potentially compelling for the programming of quantum computers, due to both the unintuitive nature of these devices as well as the unique challenges presented by near-term hardware. 

To this end, we explore reinforcement learning for automated program synthesis in the context of a hybrid classical-quantum computing architecture, considering both simulated and physical gate-based quantum computers. We explore the application of this framework to solve a range of combinatorial optimization problems (COPs). This class of problems is of particular interest in the quantum computing domain due to the emergence of new quantum heuristics for both adiabatic \cite{das2005quantum} and gate-model \cite{hadfield2019quantum,farhi2014quantum} quantum computing. From an application perspective, these optimization problems are ubiquitous and of very high value to many processes in industry. 

Reinforcement learning techniques applied to quantum computation have found success in a range of contexts, including quantum error correction \cite{fosel2018reinforcement}, as well as noisy control for gate design \cite{an2019deep,niu2019universal} and state preparation \cite{bukov2018reinforcement1,bukov2018reinforcement2,august2018taking,albarran2018measurement,zhang2019reinforcement}. Here we utilize reinforcement learning directly at the application level to solve COP programming tasks.

Combinatorial optimization has proven to be a popular target domain for machine learning methods. This work dates back to at least the last machine learning cycle of the 1980s and 1990s, where Hopfield networks were used to model a variety of problem types \cite{smith1999neural}. More recently, state of the art techniques such as recurrent encoder/decoder networks \cite{vinyals2015pointer,bello2016neural}, graph embeddings \cite{khalil2017learning}, and attention mechanisms \cite{nazari2018reinforcement} have been used to solve a range of COPs. Note that, in these applications, machine learning has been used either end-to-end to map directly from a problem instance to a solution, or alternatively as a subroutine of an already existing heuristic. For a comprehensive discussion of this intersection of domains, see Ref. \cite{bengio2018machine}.

In what follows, we detail the design considerations involved in defining an effective reinforcement learning environment, with particular emphasis on the definition of the state space, action space, and reward function. We subsequently apply this framework to train reinforcement learning agents to generate quantum programs solving combinatorial optimization problems on both simulated and real quantum resources. Relative to a test set of randomly generated problem instances, we observe that the mean performance of the trained agents exceeds the performance of both untrained agents as well as that of the leading near-term hybrid quantum algorithm typically used to solve combinatorial optimization problems, the quantum approximate optimization algorithm (QAOA) \cite{farhi2014quantum}. Following this, we briefly analyze and discuss agent strategy, consider limitations of the current work, and note promising avenues forward. 

\section{Learning environment}
In this section we broadly identify the aspects of the hybrid classical/quantum environment relevant to reinforcement learning. In particular, we indicate the state space and observations of it, the available actions, the reward, and the learning agent. These specifications have been implemented using the standard environment API proposed by OpenAI Gym \cite{brockman2016openai}. This implementation was used to conduct all experiments contained in this study.

At a high level, the framework we propose is as follows. A reinforcement learning agent, executing on a classical computing resource, incrementally produces a quantum program for execution on a quantum resource, with the goal of preparing a quantum state which serves to solve some posed problem. In the examples we consider, the problem may be described by the specification of a `problem instance' (e.g. a weighted graph) and a `reward function' (detailed below). 

\begin{figure}[ht] 
\centering
\includegraphics[width=0.5\textwidth]{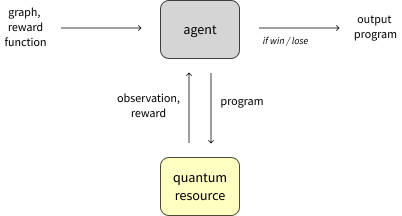}
\caption{Interaction of a learning agent with a quantum resource for the automated generation of reward-specific quantum programs. The quantum resource may be either a quantum simulation or a physical quantum processor. The win/lose criteria must be defined in the environment.}
\label{fig:interface}
\end{figure}

This process is illustrated in Fig.~\ref{fig:interface}. At the outset, the problem reward function, and graph to be evaluated in the context of this reward function are specified. Given this information, the agent produces a quantum program by iterating the following steps: i) the agent chooses from some available quantum gates to append to the current program, ii) the updated program is evaluated on a quantum resource, perhaps several times, and iii) the results of these evaluations are used to compute a reward and an `observation'. The results of step (iii) are subsequently used in the decision criteria of step (i) of the next iteration, and so on. This interaction between the agent and quantum resource is repeated until the agent `wins', wherein the reward exceeds some threshold, or `loses', wherein the program length exceeds some threshold.

In the case of combinatorial optimization problems, one may naturally identify the reward of a quantum state $\vert\psi\rangle$ as the cost Hamiltonian's expectation value. More generally, any monotonic function of the expectation serves as an adequate surrogate. For the experiments which follow, we have found it convenient to rescale the cost Hamiltonian's expectation to take values in the range 0 to 1. We identify the action space $\mathcal{A}$ as a finite set of quantum gates, such as a discretized set of RZ and RY rotation gates. For the agent, we focus exclusively on the PPO (Proximal Policy Optimization, cf. \cite{schulman2017proximal}) algorithm, applied to a shared actor-critic architecture.

A reinforcement learning problem is formally specified as a Markov Decision Process (MDP), for which the goal of the learning agent is to find the optimal policy, i.e. the conditional probability $\pi^* (a | s)$ of applying a particular quantum gate (action $a$) given a particular representation of the qubit register (state $s$) that would maximize the expected (discounted) return $\mathbb{E}_{\pi}[\sum_{k=0}^{\infty} \gamma^{k} r_{k+1}]$, without necessarily having a model of the environment $p(s^{\prime}, r \vert s, a)$. Defining the \textit{value} of a state $s$ under a policy $\pi$ as

\begin{equation}
V_{\pi}(s) = \mathbb{E}_{\pi}\left[\left.\sum_{k=0}^{\infty} \gamma^{k} r_{t+k+1} \right\vert  S_t = s \right]
\end{equation}

we identify the optimal policy a bit more concretely as $\pi^*$ such that $V_{\pi^*}(s) \geq V_{\pi}(s)$ for all $s \in \mathcal{S}$ and all policies $\pi$. In practice PPO will find some approximation to the theoretical optimum as a function of some parameters $\pi^* (a | s; \theta) \approx \pi^* (a | s)$ which it will tune towards the optimum during the learning process.

The problem we consider here is more naturally modeled as a \textit{partially observed Markov Decision Process} (POMDP), since quantum states are not themselves directly observable, and only their measurement outcomes are. While the action (quantum gate) that the agent chooses to carry out deterministically evolves the quantum state (in the absence of noise), the observation it receives from the measurement samples are in general not deterministic. For a single COP instance, the observation representation that the agent receives from the environment is given by a specification of the sampled bitstrings we receive from $N_{shots}$ measurements, given a particular sequence of actions and a fixed starting state

\begin{equation}
s_{samples} : \vert \psi \rangle \rightarrow \left[N_0, ..., N_{2^n - 1} \right]
\end{equation}

such that $\sum_{i=0}^{2^n - 1} N_i = N_{shots}$. Note that $N_i / N_{shots} \approx \vert \alpha_i \vert^2$, where $\alpha_i$ describes the complex amplitude of the computational basis state $\vert i \rangle$.

In order to extract quantum circuits from the trained agent on unseen problem instances of a COP, we augment the state space with a representation of the COP problem instance itself. For example, in the case of \textsc{MaxCut} (see below), part of the state description is the graph whose maximum cut we seek. We train the agent over a collection of several such COP instances, forming the training set, and test its predictions against a collection of similar but different COP instances that the agent has not seen before.

\section{Quantum heuristic for combinatorial optimization}

In this section, we first outline the classes of combinatorial optimization problems which we consider. Following this, we describe several experiments on simulated and real quantum hardware.

\subsection{Surveyed problems}
We consider three problems of increasing generality.

Let $G = (V,E)$ be a weighted graph with vertices $V$ and edges $E$. For convenience we assume $V = \{1,\ldots, n\}$, with weights specified by an $n \times n$ real symmetric matrix $w$, where the weight $w_{ij} \geq 0$ is nonzero if there is an edge between vertices $i$ and $j$. The maximum cut problem asks for a partition of $V$ into two subsets such that the total edge weight between them is maximized. Formally,

\leqnomode
\begin{equation}
    \tag*{\textsc{MaxCut}} \label{maxcut}
    \begin{aligned}
        & \underset{z \in \{{-1},1\}^n}{\text{maximize}} & & \frac{1}{2}\sum_{i, j} w_{ij} \frac{1 - z_i z_j}{2}.
    \end{aligned}
\end{equation}

This problem is known to be NP-hard \cite{karp1972reducibility}, although a number of polynomial-time approximation algorithms exist. In this regard, it is known to be NP-hard to approximate \textsc{MaxCut} with ratio better than 16/17 (\cite{arora1998proof}). The best known approximation ratio of 0.87856 is given by the semidefinite programming algorithm of Goemans and Williamson \cite{goemans1995improved}. If the Unique Games Conjecture holds, this ratio is optimal \cite{khot2007optimal}.

Note that solving \ref{maxcut} is equivalent to maximizing the slightly simplified expression $\sum_{i < j} (-w_{ij}) z_i z_j$, where the coefficients $(-w_{ij})$ are always negative. A natural generalization is to allow mixed signs. The resulting problem, also NP-hard, is

\begin{equation*}
    \tag*{\textsc{MaxQP}} \label{maxqp}
    \begin{aligned}
        & \underset{z \in \{{-1},1\}^n}{\text{maximize}} & & \sum_{i,j} w_{ij} z_i z_j
    \end{aligned}
\end{equation*}
where $w$ is a real symmetric matrix with null diagonal entries.

It can be shown that the optimal value in \textsc{MaxQP} is non-negative, and a randomized $\Omega(1/\log n)$-approximation algorithm is given in \cite{charikar2004maximizing} (see also \cite{nesterov1998global}, \cite{nemirovski1999maximization}, \cite{megretski2001relaxations}). It is NP-hard to approximate with a ratio better than 11/13, and quasi-NP-hard to approximate with a ratio better than $\Theta(1 / \log^{\gamma} n)$ for a constant $\gamma > 0$ \cite{arora2005non}.

Although of great theoretical interest, \textsc{MaxCut} and \textsc{MaxQP} do not necessarily offer the most convenient form in which to embed practical problems. In this regard, one may consider a slight generalization of \ref{maxqp} where the quadratic is augmented with an affine term. In keeping with the conventions of the literature, we pose this as the \textsc{QUBO} ("quadratic unconstrained binary optimization") problem, given by

\begin{equation}
    \tag*{\textsc{QUBO}} \label{qubo}
    \begin{aligned}
        & \underset{x \in \{0,1\}^n}{\text{maximize}} & & \sum_{i,j} w_{ij} x_i x_j
    \end{aligned}
\end{equation}
\reqnomode

where $w$ is a real symmetric $n \times n$ matrix. The above formulation is sometimes abbreviated as UQBP (``unconstrained quadratic binary program"), see \cite{kochenberger2014unconstrained} for a broad survey, and \cite{dunning2018works} for a recent study of the empirical performance of various heuristic algorithms.We remark that under a transformation $z_i = 1 - 2 x_i$ we may embed instances of \textsc{MaxCut} and \textsc{MaxQP} as instances of \textsc{QUBO}.

In the context of quantum computing it is standard to formulate the above optimization problems in the language of operators and expectation values, so that task is to maximize the expectation of a certain `problem Hamiltonian' with respect to a $n$-qubit quantum state. When the cost function is expressed as a polynomial in $\{-1,+1\}$ variables (e.g. as in $\sum_{i,j} w_{ij} z_i z_j$), the corresponding Hamiltonian follows immediately (e.g. $\sum_{i,j} w_{ij} Z_i \otimes Z_j$ where $Z_i \otimes Z_j$ denotes the tensor product of Pauli $Z$-operators acting on the $i$th and $j$th qubits respectively).


\subsection{Experiments} \label{experiments}
Having fixed the identification of the state, observation, action space, reward, and learning agent, we apply the aforementioned framework in order to solve random instances of the above three problems. In all cases we take the number of variables $n=10$ fixed. For \textsc{MaxCut}, we consider Erdős-Renyi graphs with edge probability $0.5$. If there is an edge from vertex $i$ to vertex $j$, we choose an independent weight $w_{ij} \sim \text{Uniform}(0,1)$. Otherwise, $w_{ij} = 0$. For \textsc{MaxQP}, we consider random matrices $w$ formed by letting $w_{ij} \sim \text{Uniform}(-1,1)$ if $i < j$, $w_{ii} = 0$, and imposing that $w$ be symmetric. For \textsc{QUBO}, we consider random matrices $w$ formed by letting $w_{ij} \sim \text{Uniform}(-1,1)$ for $i \leq j$, and imposing that $w$ is symmetric. For each of the three problems, we generated 50,000 random instances, which were then shuffled into a single dataset for training. Additional independent validation and test sets were created for each problem type. In total, the validation set contained 12000 unique instances, while the test set contained 3000 unique instances.

Following a modest amount of hyperparameter optimization (further detailed in the supplementary information), we trained a PPO agent for approximately 1,700,000 episodes (for a total of approximately 20 million individual steps) on a quantum simulator (Rigetti's ``Quantum Virtual Machine", henceforth abbreviated as QVM \cite{smith2016practical}). At the end of the training, we selected the model with the best performance on the validation set, as indicated by the highest average episode score (see below). We refer to this model as the `QVM-trained' agent. The QVM-trained agent was then used to initialize training of the `QPU-trained` agent on the Rigetti Aspen QPU \cite{didier2018ac,caldwell2018parametrically,nersisyan2019manufacturing}. This model was trained for 150,000 episodes.

We also run $p=1$ (single-step) QAOA\cite{farhi2014quantum} on each of the test sets as a benchmark (see the supplementary information for more information). We choose QAOA since it is a widely known and fairly generic algorithm for solving combinatorial optimization problems on quantum computers. Although QAOA generally performs better the larger the value of $p$, or the number of alternating steps in the algorithm, we focus on $p=1$ since for the 10-variable problems under consideration this limits the program length to approximately 100. A larger number of steps would cause the output to have significant contributions from noise, which we would like to avoid as a comparative benchmark. We additionally benchmark performance with respect to an `untrained' agent. This agent is subject to the same environment as the trained agent, but has not been optimized through a training process. This means that, for all possible observations, the untrained agent samples from the action set with uniform probability. 

With respect to a fixed problem instance and a fixed policy, one may produce an episode, described by a sequence $s_1, a_1, r_1, \ldots, s_{m}, a_m, r_m$ of state-action-reward triplets. The reward for a single action is formally the expectation value of the problem Hamiltonian in the state produced by the program sequence thus far, normalized to be in the range 0 to 1, i.e. $r_i = (\langle \psi_i \vert H_C \vert \psi_i \rangle - m)/(M-m)$ where $\vert \psi_i \rangle$ is the state prepared by the first $i$ actions, and  $m = \text{min}_{\psi} \langle \psi \vert H_C \vert \psi \rangle$ and $M = \text{max}_{\psi} \langle \psi \vert H_C \vert \psi \rangle$ denote the minimum and maximum attainable expectation values. We remark that this normalization is convenient for our subsequent data analysis, and enables the possibility of early stopping in training, but is not strictly speaking necessary for the use of our proposed method (and rightfully so, as the maximum value used in the normalization is precisely what we set out to compute). On a physical device, the reward $r_i$ is estimated by repeated preparation and measurement for some number of measurement shots (cf. the supplementary information).

\begin{figure}[H]
    \centering
    \includegraphics[width=\textwidth]{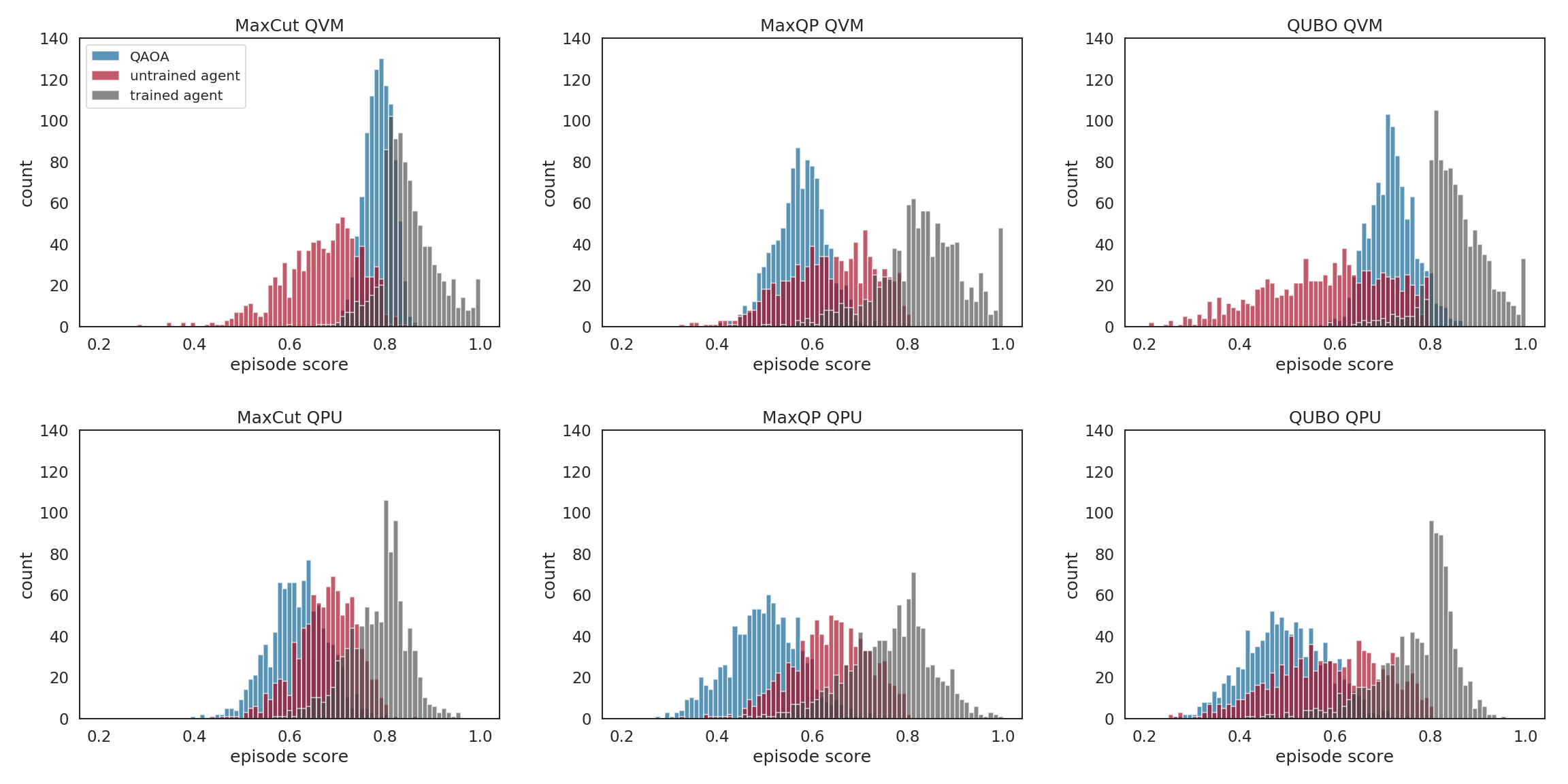}
    \caption{Distribution of episode scores, with respect to the test dataset, for each COP on the quantum simulator (top), and on the quantum processor (bottom), and for the QVM-trained agent (dark gray), the untrained agent (red), and the QAOA (blue). Overall, the QVM-trained agent yields the highest expected test performance for all problems on both the simulated and physical quantum resources.}
    \label{fig:test_reward}
\end{figure}

\begin{figure}[H]
    \centering
    \includegraphics[width=\textwidth]{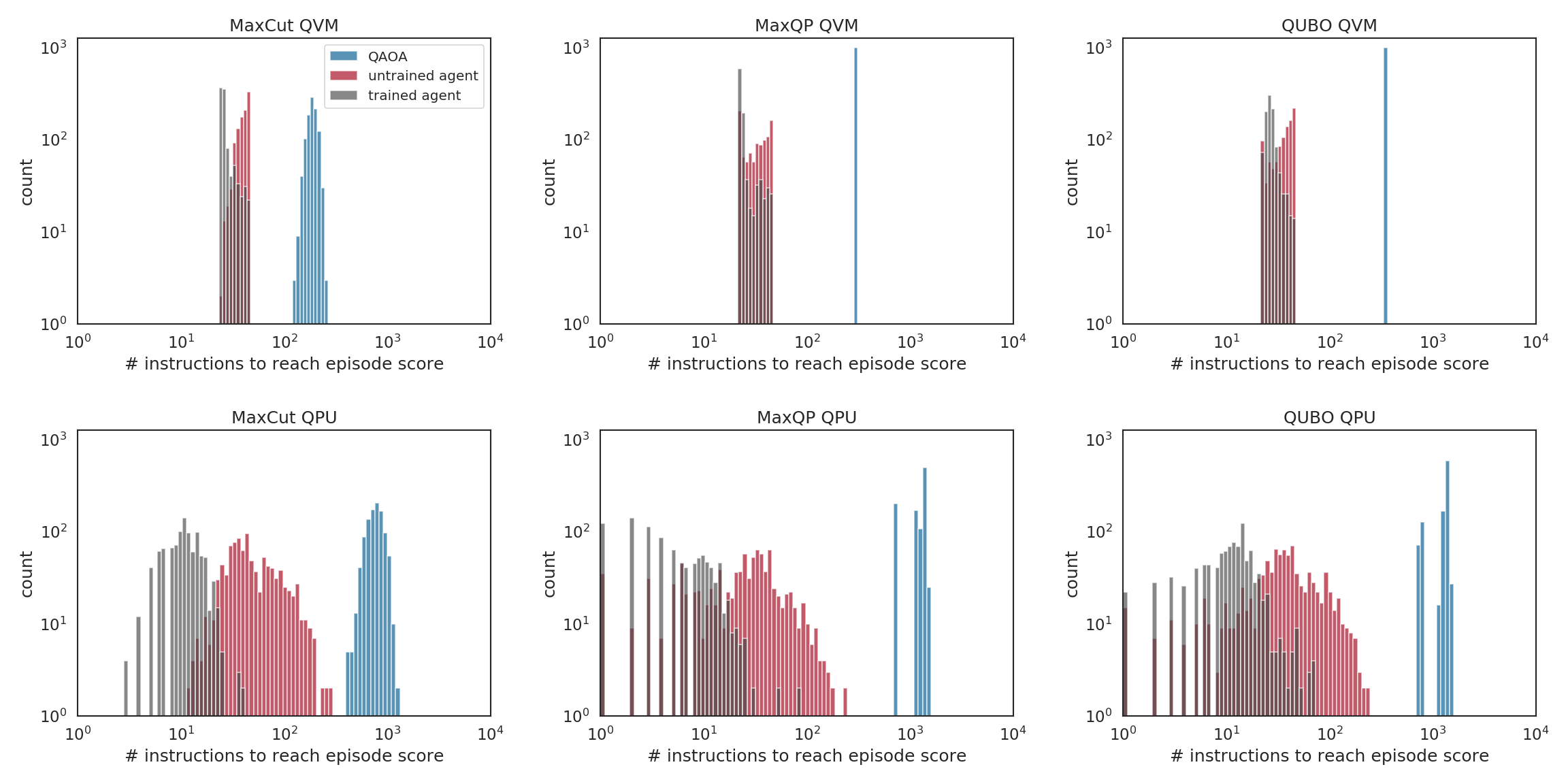}
    \caption{Program length corresponding to the episode scores given in Fig.~\ref{fig:test_reward} for each COP on the quantum simulator (top), and on the quantum processor (bottom) for the QVM-trained agent (dark gray), the untrained agent (red), and the QAOA (blue). Through training, on both resources, shorter programs are generated by the agent.}
    \label{fig:test_length}
\end{figure}

\begin{figure}[ht]
    \centering
    \includegraphics[width=\textwidth]{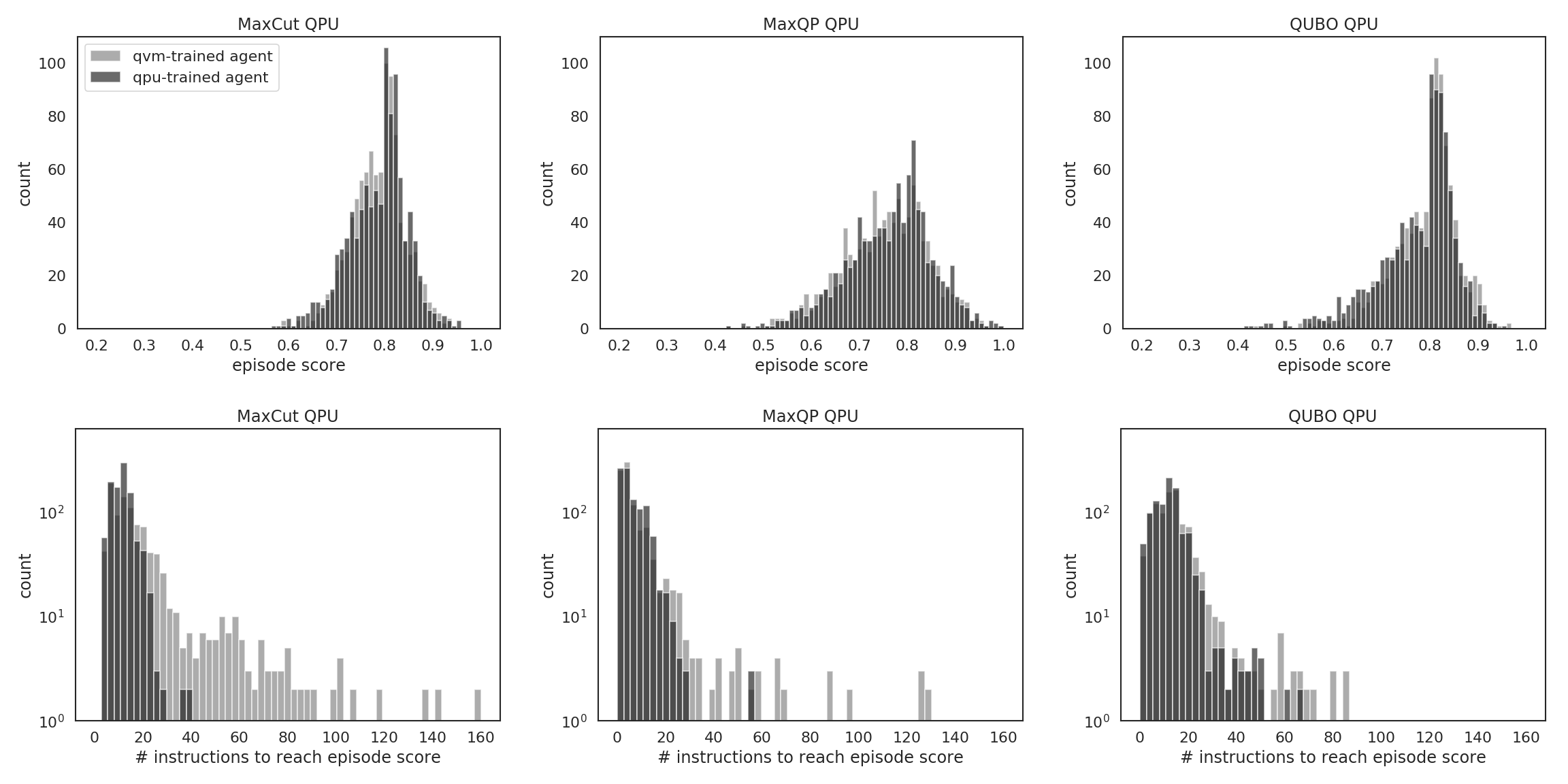}
    \caption{Test performance for each COP for both the QVM-trained agent (light gray) and the QPU-trained agent (dark gray)  the quantum processor (top), and the corresponding number of compiled instructions to reach each test episode score, for each agent, on the quantum processor (bottom). Through training on the QPU, the test score distributions yield similar empirical characteristics, showing comparable expectation and variance values. However, the number of instructions to reach these scores decreased over the course of the QPU training process.}
    \label{fig:qpu_training}
\end{figure}

With respect to a complete episode, the {\it episode score} metric is given by ES$_{\text{agent}}$ $\coloneqq \max_{1 \leq i \leq m} r_i$. For the QAOA ansatz, which is intended to run to completion, the QAOA {\it episode score} is given by ES$_{\text{QAOA}}$ $\coloneqq r_m$, where $m$ represents the final ansatz instruction.

Comparison of the performance on the test problems, for each COP, for each surveyed method, on each quantum resource is illustrated in Fig.~\ref{fig:test_reward}. In each circumstance, the QVM-trained agent produces a score distribution with the highest expected episode score. In moving from the QVM to the QPU, episode score distributions were found to increase in variance, and in the case of the QAOA, decrease in expected value. For the QVM-trained and untrained agents, expected test performance remains comparable between the QVM and QPU. Increased performance of the trained agent compared to the untrained agent, in all circumstances, suggests that the training process is successful in biasing the action distribution to yield more promising, observation-dependent, actions. Increased performance of the trained agent compared to the QAOA, in all circumstances, suggests that the $p=1$ QAOA ansatz is a less optimal circuit ansatz compared to those generated by the trained agent. Additionally, we note that the trained agent is less sensitive to COP-type compared to the QAOA, resulting in test distributions of similar character for each surveyed problem. Interestingly, the untrained agent yields better performance on the QPU compared to the QAOA. We speculate that this is in part due to the structure, length, and noise sensitivity of the QAOA ansatz.

Given the {\it episode score} test distributions given in Fig.~\ref{fig:test_reward}, one may compute the number of instructions contained within the programs generating each {\it episode score}. We label this instruction-metric {\it \# instructions to reach episode score}, and depict this data in Fig.~\ref{fig:test_length}. Note that when executing on the QVM, there is no noise, and the program is not compiled. The QVM instruction-metric therefore describes, for each {\it episode score}, the number of corresponding (perfect) gates specified by the agent (in other words, $\mbox{argmax}_{1 \leq i \leq m} r_i$). In the case of the QPU, the agent program is compiled to the chip topology of the Rigetti Aspen QPU, and the native gate set given by CZ, RZ($\theta$), and RX( $\pm \pi / 2$) using the Rigetti Quilc compiler. The QPU instruction-metric therefore describes, for each \textit{episode score}, the number of compiled gates resulting from a sequence of actions specified by the agent.

By design, the agents were constrained to 25 uncompiled instructions. For the QAOA, the uncompiled program length is determined by the size and connectivity of the graph of interest to the $p=1$ QAOA ansatz. Through training, for all problems, the expected length of the trained agent programs is decreased compared to those of the untrained agent. Although compiled, the number of instructions required to reach each episode score is reduced for the trained agent on the QPU compared to the QVM. Note that on the QVM, there is no noise. Additionally, there is no penalty for generating longer programs until the program length constraint of 25 uncompiled gates. On the QPU, the agent experiences noise that likely increases as the program increases in length and complexity. This could create an incentive for the agent to generate shorter programs on the QPU, relative to the QVM. In all cases, the untrained and trained agents yield shorter programs compared to the QAOA. On the QPU, the QAOA programs compiled into more than $10^2$ instructions. Additionally, the later gates of the QAOA are given by the more expensive entangling gates. If too much decoherence is experienced by the later stages of the program, one could expect a strong degradation in performance, as evidenced by the decreased episode score between the QVM and QPU for the QAOA.

Recall that the QPU-trained model was initialized using the parameters of the QVM-trained model, and subsequently trained for an additional 150,000 episodes on the QPU. In order to investigate the impact of training on a real quantum resource, both the QVM-trained model and QPU-trained model were test on the QPU across both the reward-metric and the instruction-metric. This is shown in Fig.~\ref{fig:qpu_training}.

Inference on both models, on the QPU, yields comparable reward-metrics. For each COP, the episode score test distributions have similar expected values and variances. Although generating comparable reward-metric statistics, the instruction-metrics look much different. It is observed that through QPU-training, the resultant instruction-metric values become much shorter. In this context, device noise acts as an indirect incentive to the agent. This indirect incentive promotes the generation of programs that remain short following compilation. This is particularly evident for the \textsc{MaxCut} problems. In this case, testing the QVM-trained model on the QPU yields compiled programs of lengths, in some cases, exceeding 100 instructions. However, testing the QPU-trained model on the QPU yields compiled programs of lengths shorter than 50 instructions.

\section{Action statistics}
\begin{figure}[ht]
    \centering
    \includegraphics[width=\textwidth]{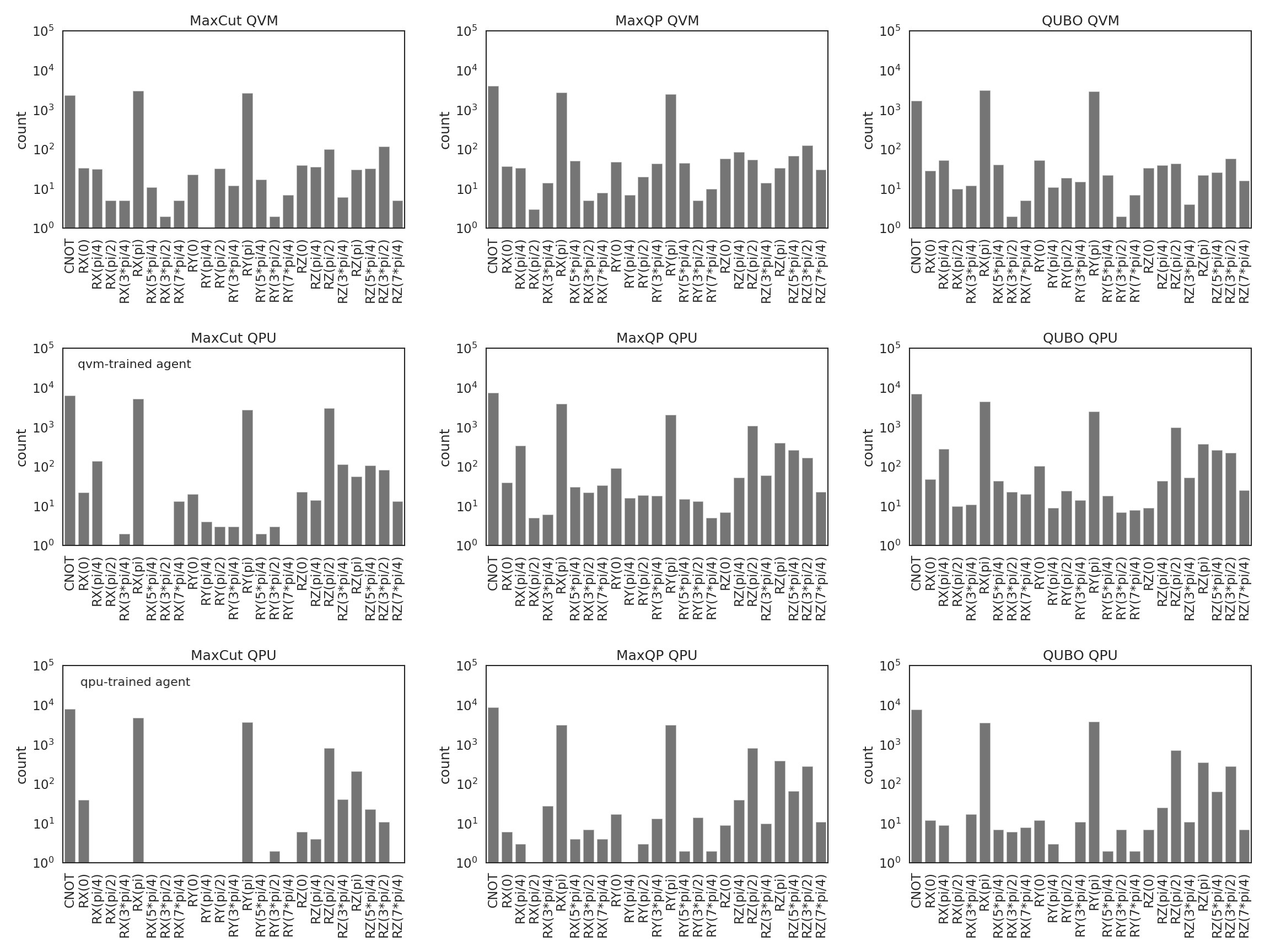}
    \caption{Action frequencies resulting from inference on the QVM with the QVM-trained model (top), and from inference on the QPU with the QVM-trained model (middle) and the QPU-trained model (bottom). In all cases, the CNOT, RX($\pi$), and RY($\pi$) gates are the most frequently implemented actions. The remaining actions are chosen with comparable frequencies. In the QPU-trained model, a reduction in the frequency of non-native gates is observed.}
    \label{fig:action_frequencies}
\end{figure}

Broad statistical analysis of agent-generated programs was accomplished by computing the frequency of each action in the programs generated by the experiments detailed in Sec.~\ref{experiments}. These histograms are shown in Fig.~\ref{fig:action_frequencies}. Note that the data between Fig.~\ref{fig:test_reward}, Fig.~\ref{fig:test_length}, Fig.~\ref{fig:qpu_training}, and Fig.~\ref{fig:action_frequencies} are all consistent, resulting from different varieties of analysis on the same set of experiments over a precomputed test dataset. Additionally, explicit examples of a set of random sampled programs are explicitly shown in the supplementary information.

The agent demonstrates a strong preference for the CNOT, RX($\pi$), and RY($\pi$) gates, occurring approximately equally for all problems, for all models, on both types of quantum resources. Interestingly, the RZ($\pi$) gate is not as consistently prevalent across the different experiments, occurring less frequently on the QVM. On the QVM, the remaining gates frequencies are distributed somewhat evenly. On the QPU, the RZ($\pi$) gate becomes more common, even in the case of the QVM-trained model. Through training on the QPU, the QPU-trained agent appears to less frequently sample non-native RY and RX gates. In contrast, the native RZ($\theta$) gates show comparable frequencies to the QVM-trained agent. This parallels the instruction-metric results shown in Fig.~\ref{fig:qpu_training}. It appears that QPU-training aims to reduce compiled program length, which amounts to diminished sampling of non-native gates (which are compiled into longer sequences of native gates). To further develop these observations, more extensive and robust analysis is required.

In the case of the COPs we have considered above with our simple reward (discounted) reward structure, we hypothesize that the theoretical limit of the optimal program would be a series of X gates. This is so because the Hamiltonians (reward functions) we consider are all diagonal in the computational basis, and their solutions can be specified as some bitstring, i.e. some computational basis element, and not necessarily a linear combination of such basis elements. The gates I and X are sufficient to produce such states starting from the $\vert 0 \rangle^{\otimes N}$ state. Hence, we expect the optimal policy to disregard any phase information in the quantum states. Furthermore, recall that our defined representation of the observation space is equivalent to the space of probabilistic bits. For example, if the goal was to maximize $\langle\psi\vert X \vert\psi\rangle$, it is sufficient to produce any of the states $(1 / \sqrt{2}) \left( \vert 0 \rangle + e^{i\theta} \vert 1 \rangle \right)$, whichever one is cheapest to produce given a particular gateset. This is another reason we therefore expect that the optimal policy should disregard any phase information. Although more extensive experimental and theoretical analysis is required, this hypothesis is consistent with the gate frequencies observed during inference on the test set (as shown in Fig.~\ref{fig:action_frequencies}).

For such problems, where the Hamiltonian is diagonal in the computational basis, the shortest sequence of gates we can imagine to produce the solution bitstring is a series of X gates on the appropriate qubits. A  rotation of any angle other than $\pi$ about the x-axis would produce a less than optimal value for the $Z_i Z_j$ term, and therefore the reward, so that we cannot use the policy improvement theorem (Ref. \cite{suttonbarto}) to improve upon this policy. 

Similarly, our choice of observation space and policy architecture reflect the correspondence between computational basis states and candidate solutions of the optimization problems. In particular, for more general problems (e.g. as may arise in quantum chemistry) it may be necessary for the observation to consist of measurements with respect to several bases; the policy in such a case should be modified accordingly. 

\section{Conclusion}
Overall we find reinforcement learning to be an effective method for quantum programming for combinatorial optimization for the problems and datasets studied here. For the evaluated problems, the trained agents were observed to generate very competitive results compared to an untrained agent and compared to $p=1$ QAOA on both simulated and physical quantum resources. 

Extensions of this work include, but are not limited to: refinements to the learning environment, training and comparison of different agent types, as well as an investigation of how the training time scales as the size of the problem grows, and application of this learning method to more programming tasks.

Modifications to the learning environment could included deeper investigation of state and observation representations. The quantum and problem observations are of a decidedly distinct nature, and thus it is natural for a policy to treat them differently. There has been much recent attention to vectorizations or feature representations of graphs and similar discrete structures, relevant to processing of a given combinatorial optimization problem. As our focus is primarily on the general method of reinforcement learning for quantum program synthesis, we have opted to not consider any specialized handling of the state observation representation and leave that as an avenue for future work. Modifications to the learning environment could also include larger or continuous action spaces, as well as modifications of the reward function. 

Additionally, we chose to focus exclusively on PPO learning agents. This is in part due to the breadth and multicomponent nature involved in iterating simultaneously on the learning environment as well as on the learning agent. The robustness of the PPO algorithm allowed for fast and favorable training without extensive hyperparameter optimization. Performance of alternative agents such as deep Q-learning is of interest.

With respect to scalability, if we limit the architecture of the shared actor-critic network to never grow more than polynomially in the size of the input problem, then the computations this network carries out in order to produce an output action should also grow at most polynomially. We may ask how many training steps we require to reach a certain validation accuracy on some held out set of problem instances. For a sufficiently high validation accuracy, the trained network may serve as a useful heuristic if the training time grows at most polynomially in the size of the problem. It is not unreasonable to expect this to happen, since neither the gate set we have considered above nor the trainable parameters of the network grow super-polynomially. However, a systematic investigation of how well this framework scales as the size of the problem input is currently lacking, and would serve as an important barometer of how well this approach would work in practice.

Here we have chosen to focus on COPs, for which the specified Hamiltonian is diagonal in the computational basis. We remain curious about the extension of this work to different domains. We speculate that extending our analysis to other problems, such as those found in quantum simulation settings where we expect to see Hamiltonians that are non-diagonal in the computational basis, would yield theoretically optimal policies that use non-Clifford operations. We also imagine that by changing the reward structure, we could retool this procedure to optimize not just for shortest sequence of gates from some given gateset, but also to preferentially utilize quantum resources over classical ones.

\section{Acknowledgements}
The authors wish to thank Amy Brown and the entire Rigetti hardware team for QPU support. We also would like to give mention to Nima Alidoust and David Rivas for their technical and logistical guidance, and Robert Smith, Joshua Combes, Eric Peterson, Marcus da Silva, Kirby Linvill, and Kung-Chuan Hsu for many insightful conversations.

\printbibliography
\end{document}


\textbf{\Large Supplementary Information: Automated quantum programming \protect via reinforcement learning \protect for combinatorial optimization}

\section{Environment and Agent Architecture} \label{sec:architecture}
Our optimization problems are particularly simple to map to quantum hardware, in the sense that we may naturally obtain a one-to-one correspondence between binary problem variables and qubits. Thus for a problem of $n = 10$ variables, each basis vector of a $10$-qubit system may be expressed in ket notation as $|b_1 \ldots b_n \rangle$ where $b_i \in \{0,1\}$, and hence a single measurement of this system in the standard basis yields a candidate solution to the optimization problem.

\subsection*{Action Space}
The \textit{actions} are 
\begin{itemize}
    \item X, Y, and Z rotations on each qubit, with discrete angles $\frac{2 \pi k}{8}$. 
    \item CNOT gates on each pair of distinct qubits.
\end{itemize}
Where $k=\{0,...,7\}$. In summary, there are 3*10*8 = 240 single qubit actions, and 45 two qubit actions. Each action may be expressed as a single instructions in the Quil ISA (\cite{smith2016practical}), e.g. \texttt{RX}$(\pi/2)$ 7. 

\subsection*{Observation Space}
Thus a sequence of actions corresponds to a Quil program. When executed on a quantum device, a corresponding quantum state is prepared. A measurement of this state with respect to the standard computational basis results in a bitstring $b = b_1 b_2 \ldots b_n$, where $b_i$ was the measured state of qubit $i$, and we have fixed $n = 10$. This process of preparation and measurement is repeated for some number of times (the ``number of shots"), resulting in a sequence of bitstrings. In what follows, we have fixed the number of shots to be $m=10$, so that the resulting {\it observation of the quantum state} is a $10 \times 10$ binary array $B = [b^{(1)}; \ldots; b^{(10)}]$. 

A problem instance is specified by a specific choice of weights $w$. For \textsc{MaxCut} and \textsc{MaxQP}, the $\binom{n}{2} = 45$ off-diagonal upper triangular entries of the weight matrix $w$ suffice to fully describe the problem instance. For \textsc{QUBO}, the $\binom{n}{2} + n = 55$ upper triangular entries are needed. These may be concisely expressed as a numeric vector $\tilde{w}$, representing an {\it observation of the problem instance}. 

The {\it observation} made by the agent consists of the joint quantum and problem observations. In other words, we consider
\begin{equation}
    obs \coloneqq (B, \tilde{w}).
\end{equation}

\subsection*{Reward}
With respect to a given problem of the form $\max_{x \in \{0,1\}^n} C(x)$, the {\it reward} associated with the observation $\texttt{obs}$ is $r(obs) = \frac{1}{10} \sum_{i=1}^{m} \frac{C(b^{(i)}) - m}{(M - m)}$, where $m = \text{min}_b C(b)$ and $M = \text{max}_b C(b)$ are the minimum and maximum attainable values of $C$. This may be seen to be an estimate of the normalized expectation of the corresponding problem Hamiltonian.

\subsection*{Win / lose criteria}
For a given episode, the agent is said to have `won' if the normalized reward was found to exceed .8. This was chosen through quick empirical experimentation. We imagine that methods such as curriculum learning could be used in order to drive the episode rewards higher. Note that by winning, the episode terminates early.

The agent is said to have `lost` if the uncompiled program length exceeds 25 instructions.

\subsection*{Policy Architecture}
Following observation, $(B, \tilde{w})$ is concatenated to a single vector and subsequently passed through two dense layers of 32 neurons each. The output is a vector of action scores. Given that the solution space for one of our combinatorial optimization problems is of size $2^{10}$, we have intentionally adopted a small and simple architecture. For larger problems, one would generally require that the number of weights in the full network scales as some small polynomial in the number of problem variables $n$.

For learning, we use the implementation of actor-critic PPO provided by \cite{stable-baselines}. The single network we described above serves as a shared actor and critic network. The weights for both the dense layers as well as those for measurement statistics, $\Theta$, were trained. 

\section{Agent training and hyperparameter optimization}
\begin{figure}[ht]
    \centering
    \includegraphics[width=\textwidth]{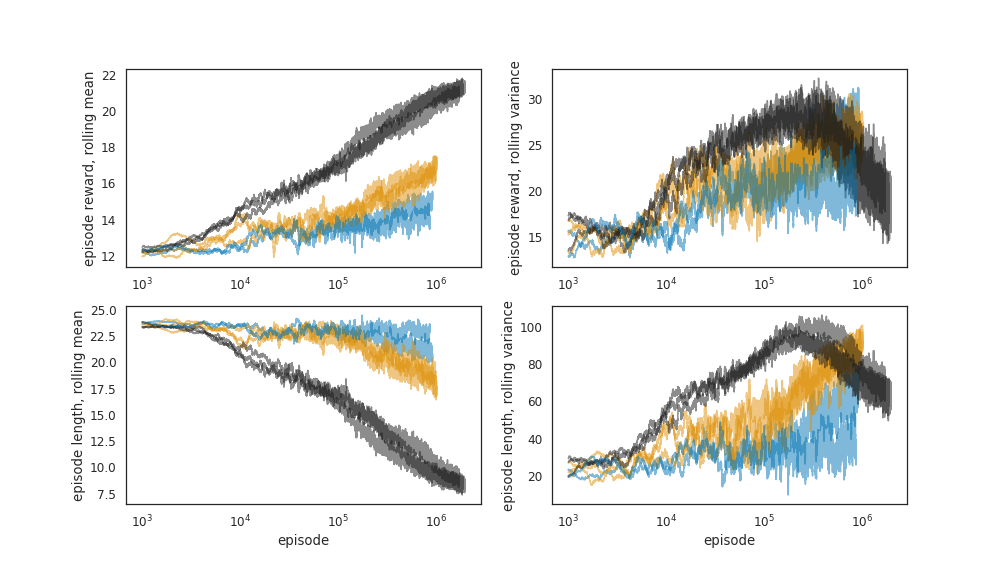}
    \caption{Learning curves for a range of PPO hyperparameter configurations.}
    \label{fig:hp_opt}
\end{figure}

PPO was found to yield reasonable results off-the-shelf, but also found to yield greatly improved results with a modest amount of hyperparameter optimization. In Fig.~\ref{fig:hp_opt} we show the learning curves for the agent under a range of agent parameterizations. It was found that increasing the $\lambda$ parameter from 0.95 (blue) to 0.999 (orange and black) resulted in improved results. Furthermore, increasing the batch size from 128 (blue, orange) to 512 (black) had a dramatic impact on agent performance. The black learning curve represent the final parameter configuration ($\lambda$ = .99, n$\_$steps = 512). 

Each curve denotes a unique random seed. It can be seen that training performance is comparable for each random seed. We note that the mean episode reward, as well as the average episode length, do not appear to have fully converged and could potentially benefit from additional training.

With the except of the hyperparameters discussed above, we used the default hyperparameters provided by the implementation in \cite{stable-baselines} (Adam optimizer with epsilon=$10^{-5}$, a linear learning schedule, clip parameter of 0.2 and advantage estimation coefficient of 0.95).

\section{Generated programs}
For illustrative purposes, we query the QVM-trained model on 8 random problems and display the corresponding output program below. Recall that these programs are iteratively constructed. So, for each program below, the final program consists of the sum of instructions in the table. $r$ denotes the evolution of the reward as the program is constructed.

\begin{tabular}{lr}
\toprule
      instr &         r \\
\midrule
   RX(pi) 8 &  0.497569 \\
   RY(pi) 1 &  0.503288 \\
   RY(pi) 0 &  0.691885 \\
   RX(pi) 9 &  0.687903 \\
 RX(pi/4) 2 &  0.676666 \\
 RX(pi/4) 2 &  0.669175 \\
 RX(pi/4) 2 &  0.661683 \\
 RX(pi/4) 2 &  0.650446 \\
 RX(pi/4) 2 &  0.654192 \\
   RY(pi) 4 &  0.713833 \\
   RX(pi) 5 &  0.723166 \\
 RX(pi/4) 2 &  0.683561 \\
   CNOT 5 9 &  0.929027 \\
\bottomrule
\end{tabular}

\begin{tabular}{lr}
\toprule
    instr &         r \\
\midrule
 RY(pi) 7 &  0.559380 \\
 RY(pi) 4 &  0.640293 \\
 RX(pi) 8 &  0.795450 \\
 RY(pi) 1 &  0.924792 \\
\bottomrule
\end{tabular}

\begin{tabular}{lr}
\toprule
        instr &         r \\
\midrule
     RX(pi) 7 &  0.734336 \\
     RX(pi) 8 &  0.710776 \\
     RX(pi) 2 &  0.742716 \\
     RY(pi) 0 &  0.751006 \\
 RY(5*pi/4) 3 &  0.837711 \\
\bottomrule
\end{tabular}

\begin{tabular}{lr}
\toprule
        instr &         r \\
\midrule
     RX(pi) 6 &  0.629431 \\
     RX(pi) 5 &  0.667715 \\
     CNOT 4 5 &  0.667715 \\
      RZ(0) 8 &  0.667715 \\
     CNOT 1 8 &  0.667715 \\
 RZ(3*pi/2) 1 &  0.667715 \\
     RX(pi) 2 &  0.760005 \\
     CNOT 5 8 &  0.904721 \\
\bottomrule
\end{tabular}

\begin{tabular}{lr}
\toprule
        instr &         r \\
\midrule
     RX(pi) 6 &  0.629431 \\
     RX(pi) 5 &  0.667715 \\
     CNOT 4 5 &  0.667715 \\
      RZ(0) 8 &  0.667715 \\
     CNOT 1 8 &  0.667715 \\
 RZ(3*pi/2) 1 &  0.667715 \\
     RX(pi) 2 &  0.760005 \\
     CNOT 5 8 &  0.904721 \\
\bottomrule
\end{tabular}

\begin{tabular}{lr}
\toprule
        instr &         r \\
\midrule
     RX(pi) 4 &  0.674805 \\
     RX(pi) 2 &  0.685028 \\
     RY(pi) 0 &  0.763762 \\
     RX(pi) 8 &  0.735677 \\
     RX(pi) 5 &  0.792516 \\
     RY(pi) 3 &  0.771576 \\
   RZ(pi/2) 8 &  0.771576 \\
 RZ(5*pi/4) 4 &  0.771576 \\
   RZ(pi/2) 8 &  0.771576 \\
   RZ(pi/2) 8 &  0.771576 \\
   RZ(pi/2) 8 &  0.771576 \\
 RZ(3*pi/2) 3 &  0.771576 \\
     CNOT 6 7 &  0.771576 \\
      RY(0) 4 &  0.771576 \\
   RZ(pi/2) 8 &  0.771576 \\
     CNOT 6 7 &  0.771576 \\
   RZ(pi/2) 8 &  0.771576 \\
     CNOT 0 9 &  0.737719 \\
     CNOT 0 9 &  0.771576 \\
   RZ(pi/2) 8 &  0.771576 \\
     CNOT 2 8 &  0.801347 \\
\bottomrule
\end{tabular}

\begin{tabular}{lr}
\toprule
      instr &         r \\
\midrule
   RY(pi) 7 &  0.787879 \\
   RX(pi) 8 &  0.712911 \\
   RY(pi) 9 &  0.712158 \\
 RZ(pi/2) 2 &  0.712158 \\
   RX(pi) 5 &  0.938972 \\
\bottomrule
\end{tabular}

\begin{tabular}{lr}
\toprule
    instr &         r \\
\midrule
 RX(pi) 4 &  0.464864 \\
 RX(pi) 2 &  0.605566 \\
 RX(pi) 8 &  0.765052 \\
 RX(pi) 9 &  0.912750 \\
\bottomrule
\end{tabular}

\section{QAOA Experiments}
QAOA with $p=1$ consists of optimizing two classical parameters, conventionally denoted $\gamma$ and $\beta$, such that the expectation value of the problem Hamiltonian is maximized by the quantum state defined by
\begin{equation}
\vert \gamma, \beta \rangle = e^{-\imath\beta B} e^{-\imath\gamma H_C} \vert + \rangle^{\otimes n}
\end{equation}
where $n$ denotes the number of qubits (10 in our case), $\vert + \rangle^{\otimes n} = H^{\otimes n} \vert 0 \rangle^n$ is the equal-superposition state of all bitstrings, $C$ denotes the problem Hamiltonian, and $B$ denotes the `mixer' which is conventionally taken to be $B=\sum_{j=1}^n X_j$.
For every problem instance, we use the \verb|Rigetti WavefunctionSimulator| to compute optimal values of $\gamma$ and $\beta$. We do this by discretizing the interval $[0, 2\pi)$ into 20 bins, and then looping over a 20 x 20 grid, corresponding to the various values of $\gamma$ and $\beta$, and calculating the expectation value $\langle \gamma, \beta \vert C \vert \gamma, \beta \rangle$. These expectation values are calculated exactly via matrix multiplication, and not estimated through sampling.

Once we know the optimal values of $\gamma$ and $\beta$ for each problem instance, we run the corresponding circuit on the actual quantum processor, and compute the solution quality as the expectation value of the cost Hamiltonian for that specific problem instance in the state produced by the $p=1$ QAOA quantum circuit, normalized by the difference between the maximum and minimum expectation value of that cost Hamiltonian. The maximum cost is computed by simply looping over all possible $2^{10}$ bitstrings, calculating the value of the cost Hamiltonian for each of those bitstrings, and reporting the maximum. Concretely, we define the solution quality $\langle \gamma, \beta \vert \tilde{C}\vert \gamma, \beta \rangle$ as

\begin{equation}
\langle \gamma, \beta \vert \tilde{C} \vert \gamma, \beta \rangle = \frac{\langle \gamma, \beta \vert C \vert \gamma, \beta \rangle - \min_{\vert \psi \rangle} \langle \psi \vert C \vert \psi \rangle}{max_{\vert \psi \rangle} \langle \psi \vert C \vert \psi \rangle - \min_{\vert \psi \rangle} \langle \psi \vert C \vert \psi \rangle}
\label{eq:soln-quality}
\end{equation}

which coincides with the definition of the reward for the RL agent. All the terms appearing on the right hand side of Eq.\ref{eq:soln-quality} are computed exactly via classical methods, except for the expectation value $\langle \gamma, \beta \vert C \vert \gamma, \beta \rangle$, which is estimated via sampled bitstrings. For each problem instance, we sample 10 bistrings to estimate this expectation value. For every problem instance, we also compute the length of the compiled $p=1$ QAOA quantum program.

\printbibliography